\documentclass{ws-jnmp}

\usepackage{amssymb, amsmath, textcomp, dsfont, mathrsfs, verbatim}

\begin{document}


\markboth{Jennie D'Ambroise and Floyd L. Williams}
{Parametric solution of certain nonlinear differential equations in cosmology}

%
%
\copyrightauthor{J. D'Ambroise and F. L. Williams}

\title{PARAMETRIC SOLUTION OF CERTAIN NONLINEAR DIFFERENTIAL EQUATIONS IN COSMOLOGY}

\author{\footnotesize JENNIE D'AMBROISE}

\address{Division of Science and Mathematics, University of Minnesota - Morris\\
Morris, Minnesota 56267, USA\\
\email{jdambroi@morris.umn.edu}}

\author{FLOYD L. WILLIAMS}

\address{Department of Mathematics and Statistics, University of Massachusetts\\
Amherst, Massachusetts 01003, USA\\
\email{williams@math.umass.edu}}

\maketitle

\begin{history}
\received{(Day Month Year)}
\revised{(Day Month Year)}
\accepted{(Day Month Year)}
\end{history}

\begin{abstract}
We obtain in terms of the Weierstrass elliptic $\wp-$function, sigma function, and zeta function an explicit parametrized solution of a particular nonlinear, ordinary differential equation.  This equation includes, in special cases, equations that occur in the study of both homogeneous and inhomogeneous cosmological models, and also in the dynamic Bose-Einstein condensates - cosmology correspondence, for example. 
\end{abstract}


\ccode{2000 Mathematics Subject Classification: 33E05, 83C20, 83F05}

\section{Introduction}	

In this paper we solve the nonlinear differential equation
\begin{equation}\dot{Y}(t)^2=\frac{f(Y(t))}{Y(t)^{2n}},\label{eq: Yeqngeneraln}\end{equation}
where $f(x)=a_0x^4+4a_1x^3+6a_2x^2+4a_3x+a_4$ is a quartic polynomial with no repeated factors and $n\geq 0$ is a fixed whole number.  The solution is expressed parametrically in terms of the Weierstrass $\wp$-function $\wp(w)$, and his sigma and zeta functions $\sigma(w), \zeta(w)$; see formulas (2.4), (2.5), (2.6), (2.7), and definitions (A.1), (A.5) of the Appendix.  One can also solve equation (1.1) in case $f(x)$ does have a repeated factor -- a situation  which is easier to deal with and which, in particular, generally does not involve elliptic functions; see Example 6.

Of some special interest is the equation
\begin{equation}\dot{Y}(t)^2=BY(t)^2+EY(t)-K+\frac{A}{Y(t)}+\frac{D}{Y(t)^2},\label{eq: Yeqn}\end{equation}
with $n=1$ in (1.1).  Note that the solution of the differential equation
\begin{equation}\dot{Y}_1(t)^2=Y_1(t)^2\left[ B_1+E_1 Y_1(t)^m - K_1 Y_1(t)^{2m}+A_1 Y_1(t)^{3m}+D_1 Y_1(t)^{4m}\right],\label{eq: Yeqnwithm}\end{equation}
where $m\neq 0$ is a fixed whole number, can be obtained  from that of (\ref{eq:  Yeqn}).  Namely, if we set $Y(t)=Y_1(t)^{-m}$ then equation (\ref{eq: Yeqnwithm}) is transformed to (\ref{eq: Yeqn}) for $B=m^2B_1, E=m^2E_1, K=m^2K_1, A=m^2A_1$, and $D=m^2D_1$.  

If both $E$ and $D$ are zero, for example, then the general solution $Y_{E,D=0}$ of (\ref{eq: Yeqn}), in terms of the Weierstrass elliptic function $\wp(w)$, was obtained in 1933 by G. Lema\^itre \cite{Lemaitre}, in his study of spherically symmetric distributions of matter.  Compare also the paper \cite{Omer} of G. Omer with references therein to special case solutions by R. Tolman, B. Datta, and H. Bondi.  The solution $Y_{E,D=0}$ provides for an exact, inhomogeneous cosmological solution $ds^2$ of the Einstein field equations with cosmological constant $\Lambda=3B$.  Namely, for the family of \emph{Szekeres-Szafron solutions}

\begin{equation}ds^2=dt^2-e^{2B(x,y,z,t)}(dx^2+dy^2)-e^{2A(x,y,z,t)}dz^2,\label{eq: SSmetric}\end{equation}
the functions $B(x,y,z,t), A(x,y,z,t)$ are explicated by the solution $Y_{E,D=0}$.  The reference \cite{KW}, for example, contains a detailed discussion of this matter, and the text \cite{InhomCosMod} can be consulted for a comprehensive analysis of inhomogeneous cosmology.  Also see \cite{Kras}, where one can consider $D=-Q^2\neq 0$ (with yet $E=0$), $Q$ being a constant electric charge.

The Friedmann-Lema\^itre-Robertson-Walker (FLRW) metric can be obtained from $ds^2$ in (\ref{eq: SSmetric}) (as a ``limit"), and a known formula (see formula (42) of \cite{KW}, for example) for the scale factor (the ``radius" of the FLRW universe) also follows from the general formulas presented here.  Further remarks on this, as well as the computation of scale factors in anisotropic models (in the Bianchi $V$ and $IX$ models, for example) are taken up in section 3.

Another case of interest, among others to be mentioned later, is that when both $A$ and $B$ are zero in (\ref{eq: Yeqn}).  Here the solution $Y_{A,B=0}$ is of relevance regarding the dynamic correspondence between Bose-Einstein condensates (BECs) and FLRW/Bianchi $I$ cosmology \cite{LidseyBEC, JDFW}.  In particular we deduce an alternate formula (in section 3) for the second moment $I_2(t)=Y_{A,B=0}(t)$ of the wavefunction of the Gross-Pitaevskii equation, for BECs governed by a time-dependent, harmonic trapping potential - especially when a cosmological constant is present - i.e. when $E\neq 0$.  The second moment in fact determines the harmonic trapping frequency.

Our formulas therefore provide for a general, unifying context where some known formulas in the literature are immediately derived or extended, and some new ones are developed -- as seen in the concrete examples of section 3.

\section{The Solution Of Equation (\ref{eq: Yeqngeneraln})}

\indent\indent For  $Z\stackrel{def.}{=} \sqrt{ a_0 Y^4+4a_1Y^3+6a_2Y^2+4a_3Y+a_4}$ equation (1.1) is expressed as 
\begin{equation}t=\displaystyle\int\frac{Y^n}{Z}dY+\delta\end{equation}
for an integration constant $\delta$.   By a choice of any root $x_0$ of $f(x_0)=0$ and a (finite) Taylor expansion $f(x)=4\alpha_3(x-x_0)+6\alpha_2(x-x_0)^2+4\alpha_1(x-x_0)^3+\alpha_0(x-x_0)^4$ of $f(x)$ about $x_0$, where $\alpha_3=f'(x_0)/4, \alpha_2=f''(x_0)/12$, $\alpha_1=f'''(x_0)/24$, $\alpha_0=f''''(x_0)/24$ one can reduce the elliptic integral $I$ in (2.1) to a \emph{Weierstrass canonical form}, where the quartic in $Z$ is reduced to a cubic.  Namely, by the substitution $Y=x_0+\alpha_3/(x-\alpha_2/2)$ one gets 
\begin{equation}I = \displaystyle\int \frac{\left( x_0+\frac{\alpha_3}{x-\alpha_2/2}\right)^n}{\sqrt{4x^3-g_2x-g_3}} dx\end{equation}
where
\begin{eqnarray}
g_2&=& 3\alpha_2^2-4\alpha_1\alpha_3=a_0a_4-4a_1a_3+3a_2^2,\\
g_3&=&2\alpha_1\alpha_2\alpha_3-\alpha_2^3-\alpha_0\alpha_3^2=a_0a_2a_4+2a_1a_2a_3-a_2^3-a_0a_3^2-a_1^2a_4\notag\end{eqnarray}
are the \emph{Weierstrass invariants} of $f(x)$.  Here $\alpha_3\neq 0$ since $f'(x)\neq 0$, as $f(x)$ has no repeated factors, by hypothesis.  Moreover if $\wp(w)=\wp(w; g_2, g_3)$ is the \emph{Weierstrass $\wp$-function} attached to $g_2, g_3$ (as in the Appendix) then, by equation (A.2) there, the substitution $x=\wp(w+c)$, for any fixed constant $c$, leads to the re-statement
\begin{equation}t=\displaystyle\int \left( x_0 + \frac{f'(x_0)}{4\left[ \wp(w+c)-f''(x_0)/24\right]}\right)^n dw+\delta\end{equation}
of equation (2.1), where we also have (by the above substitution $Y\rightarrow x$)
\begin{equation} Y=x_0+\frac{f'(x_0)}{4\left[ \wp(w+c)-f''(x_0)/24\right]}.\end{equation}

Equations (2.4), (2.5), which constitute the main result, provide for a parametric solution of equation (1.1).  For the applications we have in mind, in examples 1-6 below, we need only the cases $n=0, 1, 2$.  For $n=0$ and the choice $\delta=0$, $t=w$ in (2.4) and formula (2.5) also follows from a general formula of Biermann-Weierstrass \cite{Biermann}; more on this point later.  Let $\sigma(w), \zeta(w)$ denote the Weierstrass \emph{sigma} and \emph{zeta functions} respectively; see definition (A.5).  For a choice $w_0$ with $\wp(w_0)=f''(x_0)/24$ we have for $n=1, 2$ \emph{respectively} in (2.4)
\begin{equation}t=x_0w + \frac{f'(x_0)}{4\wp'(w_0)}\left[ \log \frac{\sigma(w+c-w_0)}{\sigma(w+c+w_0)} + 2(w+c)\zeta(w_0) \right] +\delta, \label{eq: n=1integration} \end{equation}
\begin{eqnarray}t&=&x_0^2w+\left[ - \frac{x_0 f'(x_0)}{2\wp'(w_0)}+\frac{f'(x_0)^2\wp''(w_0)}{16 \wp'(w_0)^3}\right] \log \frac{\sigma(w+c+w_0)}{\sigma(w+c-w_0)}\notag\\
&& - \frac{f'(x_0)^2}{16\wp'(w_0)^2}\left[ \zeta(w+c+w_0)+\zeta(w+c-w_0)\right]\label{eq: n=2integration}\\
&& + (w+c)\left( \frac{x_0 f'(x_0)}{\wp'(w_0)}\zeta(w_0)-\frac{f'(x_0)^2}{16}\left[ \frac{2\wp(w_0)}{\wp'(w_0)^2}+\frac{2\wp''(w_0)\zeta(w_0)}{\wp'(w_0)^3}\right] \right) + \delta,\notag
\end{eqnarray}
by formulas 1037.06, 1037.11, respectively, in \cite{BF} -- assuming also that $\wp(w_0)\neq$ the roots $e_1, e_2, e_3$ of $4x^3-g_2x-g_3=0$.  In fact, in principle, one can compute the integral in (2.4) for \emph{arbitrary} $n\geq 0$, using formulas in \cite{BF}, which we shall not pursue as the result is not needed here.

The aforementioned Biermann-Weierstrass formula for the solution $Y(t)$ of equation (1.1) in case $n=0$ is
\begin{equation}Y(t)=Y(0)+ \frac{  \left[ f(Y(0))^{1/2}\wp'(t)+\frac{f'(Y(0))}{2}\left( \wp(t)-\frac{f''(Y(0))}{24}\right)+\frac{f(Y(0))f'''(Y(0))}{24}\right]}{2\left[ \wp(t)-\frac{f''(Y(0))}{24}\right]^2-\frac{f(Y(0))f''''(Y(0))}{48}}.\end{equation}
Also see references \cite{Reynolds, WW}.

\section{Some examples}

The following examples are meant to provide application and further clarity of the preceding formulas.

\begin{example}
We begin with equation (20)
\begin{equation}\dot\Phi(t)^2=-K(z)+\frac{2M(z)}{\Phi(t)}+\frac{\Lambda \Phi(t)^2}{3}\label{eq: SSeqn}\end{equation}
of \cite{KW}, which is associated with the metric (\ref{eq: SSmetric}) in the introduction.  Since the functions $K(z), M(z)$ are independent of $t$, equation (\ref{eq: SSeqn}) is mathematically the same as equation (8.9) of \cite{Lemaitre}, and since its solution is presented and  discussed in \cite{KW} we greatly limit our remarks here.  (\ref{eq: SSeqn}) is equation (\ref{eq: Yeqn}) with $3B=\Lambda$ (a cosmological constant), $E=0, K=K(z), A=2M(z), D=0$.  Thus $f(x)=\left( \Lambda/3\right) x^4 -K(z)x^2+2M(z)x$ has $x_0=0$ as a first order root for $M(z)\neq 0$, since $f'(0)=2M(z)$.  By formulas (2.5), (2.6) the solution of (\ref{eq: SSeqn}) is given parametrically by
\begin{eqnarray}
\Phi&=& \frac{M(z)/2}{\wp(w+c; g_2, g_3)+K(z)/12}\notag\\
&&\label{eq: SSparamsoln}\\
t&=&\frac{M(z)}{2\wp'(w_0)}\left[ \log \frac{\sigma(w+c-w_0)}{\sigma(w+c+w_0)}+2(w+c)\zeta(w_0)\right]+\delta\notag\end{eqnarray}
for $\wp(w_0)=-K(z)/12\neq e_1, e_2, e_3$, which are formulas (21), (23) of \cite{KW}, where $g_2=K(z)^2/12, g_3=K(z)^3/216-\Lambda M(z)^2/12$ (by (2.3)),  and where now the integration constant $\delta=\delta(z)$ depends on the variable $z$.

\end{example}

\begin{example}
Consider the Friedmann equation
\begin{equation}a'(\eta)^2=-\kappa a(\eta)^2+\frac{K A_r}{c^4}+\frac{KA_m}{c^3}a(\eta)+\frac{c^2\Lambda}{3}a(\eta)^4\label{eq: EFEconformaltime}\end{equation}
for the scale factor $a(\eta)$ in a FLRW universe whose energy and matter are modeled by a perfect fluid.  Here $\eta$ is \emph{conformal time}, $\kappa$ is the curvature parameter, $K=8\pi G/3$ for the Newton constant $G$, $A_r$ and $A_m$ are radiation and matter constants, $c$ is the speed of light, and $\Lambda$ is a cosmological constant.  Given the ``big bang" initial condition $a(0)=0$ one derives immediately from (2.8) the result
\begin{equation}a(\eta)=\frac{\frac{\left(KA_r\right)^{1/2}}{c^2}\wp'(\eta)+\frac{KA_m}{2c^3}\left( \wp(\eta)+\kappa/12\right)}{2\left(\wp(\eta)+\kappa/12\right)^2-K\Lambda A_r/6c^2},\label{eq: aetasolution}\end{equation}
with known application in the study of cosmic microwave background fluctuation, for example; see \cite{AS, AST}.  The invariants $g_2, g_3$ in (2.3) are given by 
\begin{eqnarray}
g_2&=& \frac{\Lambda K A_r}{3c^2}+\frac{\kappa^2}{12},\notag\\
&&\\
g_3&=&-\frac{\kappa \Lambda K A_r}{18c^2}+\frac{\kappa^3}{216}-\frac{K^2A_m^2\Lambda}{48c^4}.\notag\end{eqnarray}
Elliptic function solutions of Friedmann equations are also discussed in \cite{JDMSRI, Kharb, ACB}, for example.
\end{example}

\begin{example}  As another example we consider the Bianchi V cosmological model with metric
\begin{equation}ds^2=-dt^2+X(t)^2dx^2+e^{2b x}Y(t)^2dy^2+e^{2b x}Z(t)^2dz^2\end{equation}
for $b\neq 0$.  As before, we take the energy momentum tensor to be that of a perfect fluid and denote the radiation and matter constants by $A_r, A_m$, respectively.  For $K=8\pi G/3$, a zero cosmological constant $\Lambda=0$, and the speed of light taken to be $c=1$, the Einstein equation is a special case of (1.1)
\begin{equation}\dot{R}(t)^2=\frac{1}{R(t)^4}\left[b^2 R(t)^4+KA_m R(t)^3+KA_r R(t)^2+KD\right]\end{equation}
in terms of $R(t)\stackrel{def.}{=}\left(X(t)Y(t)Z(t)\right)^{1/3}$ and the quantity $D\stackrel{def.}{=}$ $\frac{R(t)^2}{9K}\left(\frac{\dot{X}^2}{X^2}+\frac{\dot{Y}^2}{Y^2}+\frac{\dot{Z}^2}{Z^2}-\frac{\dot{X}\dot{Y}}{XY}\right.$ $\left.-\frac{\dot{X}\dot{Z}}{XZ}-\frac{\dot{Y}\dot{Z}}{YZ}\right)$ which can be shown to be a constant; see \cite{JDthesis}.    Then by (2.5) and (2.7) we obtain 
\begin{eqnarray}
R&=&x_0+\frac{f'(x_0)}{4\left[\wp(w+c)-f''(x_0)/24\right]}\nonumber\\
&&\\
t&=&x_0^2 w+\left( - \frac{x_0 f'(x_0)}{2\wp'(w_0)}+\frac{f'(x_0)^2\wp''(w_0)}{16\wp'(w_0)^3}\right) \log \frac{\sigma(w+c+w_0)}{\sigma(w+c-w_0)}\notag\\
&& -\frac{f'(x_0)^2}{16\wp'(w_0)^2}\left[ \zeta(w+c+w_0)+\zeta(w+c-w_0)\right] \notag\\
&&+(w+c)\left( \frac{x_0f'(x_0)}{\wp'(w_0)}\zeta(w_0)-\frac{f'(x_0)^2}{16}\left[\frac{2\wp(w_0)}{\wp'(w_0)^2}+\frac{2\wp''(w_0)\zeta(w_0)}{\wp'(w_0)^3}\right]\right)+\delta\notag
\end{eqnarray}
for $\wp(w_0)\stackrel{def.}{=}f''(x_0)/24$ and associated polynomial $f(x)=b^2 x^4+KA_m x^3+KA_r x^2+KD$.  The invariants are
\begin{eqnarray}
g_2&=&K\left(b^2D+\frac{A_r^2 K}{12}\right) \notag\\
&&\\
\notag g_3&=&\frac{K^2}{6}\left(b^2A_r D-\frac{A_r^3K}{36}-\frac{3DA_m^2K}{8}  \right),\end{eqnarray}
where as in the Appendix we assume that $g_2^3-27g_3^2\neq 0$.  To obtain the metric one writes $X=R, Y=Re^{\sqrt{3DK}\tau}, Z=Re^{-\sqrt{3DK}\tau}$ where $\tau(t)=\int dt/{R(t)^3}$.  That is, $\tau$ is given parametrically in terms of $w$ by
\begin{equation}\tau=
\frac{w}{x_0}-\frac{f'(x_0)}{4x_0^2\wp'(w_1)}\left[ \log\frac{\sigma(w+c-w_1)}{\sigma(w+c+w_1)}+2(w+c)\zeta(w_1)\right] + \delta' 
\end{equation}
for integration constant $\delta'$, $\wp(w_1)\stackrel{def.}{=}f''(x_0)/2-f'(x_0)/4x_0$ and assuming $x_0\neq 0$.

\end{example}

\begin{example}
For a cosmological model with Bianchi IX metric
\begin{equation}ds^2=-dt^2+a(t)^2dx^2+b(t)^2dy^2+\left[ b(t)^2\sin^2 y+a(t)^2\cos^2 y\right]dz^2-2a(t)^2\cos y \  dxdz,\label{eq: BIXmetric}\end{equation}
and massless scalar field $\Phi$ and flat (constant) potential $V(r)=2\lambda$, the modified Einstein equations based on Lyra geometry are studied in \cite{BBDR}, for example.  Also compare the paper \cite{BCHKR}.  The field equations yield the relation $\dot\Phi=\Phi_0/ab^2$ for an integration constant $\Phi_0$, and the assumption $a=b^n$ leads to the Einstein equation 
\begin{equation}\frac{\ddot{b}}{b}+(n+1)\frac{\dot{b}^2}{b^2}=\frac{1}{(n-1)b^2}-\frac{b^{2n-4}}{n-1}\label{eq: BIXbeqn}\end{equation}
for $n\neq 1$, which moreover is shown to have the first integral 
\begin{equation}\dot{b}^2=\frac{1}{n^2-1}-\frac{b^{2n-2}}{2n^2-2n}+D_1b^{-2n-2}\label{eq: BIXbeqnintegrated}\end{equation}
for $n\neq 0, \pm 1$, where $D_1$ is an integration constant.  The authors found solutions of (\ref{eq: BIXbeqnintegrated}) only for $D_1=0$ - for $n=2, 1/2, 3/2, 3/4$.  Therefore we consider the case $D_1\neq 0$, and we take $n=2$ for example:  $\dot{b}^2=1/3-b^2/4+D_1/b^6$, or $a=b^2 \Rightarrow$
\begin{equation}\dot{a}(t)^2=\frac{4}{3}a(t)-a(t)^2+\frac{D}{a(t)^2},\label{eq: BIXaeqn}\end{equation}
which is another example of equation (\ref{eq: Yeqn}) for $D=4D_1, A=0, K=0, E=\frac{4}{3}, B=-1.$  Here
\begin{eqnarray}
f(x)&=&-x^4+\frac{4}{3}x^3+4D_1,\label{eq: BIXinvars}\\
g_2&=&-4D_1, g_3=-\frac{4D_1}{9}=\frac{g_2}{9}.\notag\end{eqnarray}
In particular $g_2^3-27g_3^2=-16D_1^2(4D_1+\frac{1}{3})\neq 0$ for $D_1\neq 0, -\frac{1}{12}$, which we assume.  Certainly if $D_1=-\frac{1}{12}$, then $f(x)=-x^4+\frac{4}{3}x^3-\frac{1}{3}=-(x-1)^2\left(x^2+\frac{2}{3}x+\frac{1}{3}\right)$ has $x_0=1$ as a repeated root, for example.

On the other hand, consider $x_0=-1$ which is a non-repeated root for $D_1=\frac{7}{12}: \ f'(x_0)=8\neq 0$.  By (2.5), (2.6)   we have the parametric solution
\begin{eqnarray}
a&=&-1+\frac{2}{\wp(w+c)+5/6},\notag\\
&&\\
t&=&-w+\frac{2}{\wp'(w_0)}\left[ \log \frac{\sigma(w+c-w_0)}{\sigma(w+c+w_0)}+2(w+c)\zeta(w_0)\right]+\delta\notag\end{eqnarray}
of equation (\ref{eq: BIXaeqn}) for $\wp(w_0)=-5/6$.  One can find solutions to equation (\ref{eq: BIXaeqn}) for many other non-zero values of $D_1$ similarly.  This amounts to specifying $D_1\neq -\frac{1}{12}$ and solving for a corresponding root $x_0$ of $f(x)$ in (\ref{eq: BIXinvars}).
\end{example}

\begin{example}
The differential equation
\begin{equation}\dot{Y}(t)^2=EY(t)-K+\frac{D}{Y(t)^2},\label{eq: BECYeqn}\end{equation}
which is another example of equation (\ref{eq: Yeqngeneraln}) (for $n=1$), arises in the study of time-dependent, harmonically trapped Bose-Einstein condensates, as indicated in the introduction.  The constant $E$ here is a positive multiple of a $d-$dimensional cosmological constant, for $d\geq 3$ arbitrary.  Also $D$ is positive and $E\neq 0$.  See equation (35) in \cite{JDFW}, where some elliptic function solutions are discussed.  
Equations (2.5), (2.6) provide for a parametric solution of (\ref{eq: BECYeqn}).  However, as $f(x)=Ex^3-Kx^2+D$ is \emph{cubic} in this case, one has an alternate, simpler parametrization which in particular does not involve a logarithm, as in (2.6).

Namely consider the simple substitution $Y=ax+b$, $a\neq 0$, a suggestion for which we thank one of the referees.  Equation (2.1) then reads
\begin{eqnarray}
t&=& \displaystyle\int \frac{(a^2 x+ab)dx}{\sqrt{Ea^3x^3+(3Ea^2b-Ka^2)x^2+(3Eab^2-2Kab)x+Eb^3-Kb^2+D}}+\delta\nonumber\\
&&\\
&=& \frac{1}{a^2}\displaystyle\int \frac{(u+ab)du}{\sqrt{4u^3-g_2 u-g_3}}+\delta,\nonumber\end{eqnarray}
for $u=a^2x, a=(E/4)^{1/3}, b=K/3E, g_2'=(-3Eb^2+2Kb)/a$, and $g_3'=-Eb^3+Kb^2-D$.  Note that $g_2', g_3'$ are \emph{not} the invariants $g_2=K^2/12, g_3=\frac{K^3}{216}-\frac{E^2D}{16}$ of $f(x)$ given in definition (2.3).  Similar to the derivation of equation (2.4), we now let $u=\wp(w)=\wp(w; g_2', g_3')$ to get the parametric solution
\begin{eqnarray}
t &=& -\frac{1}{a^2}\zeta(w)+\frac{b}{a}w+\delta\nonumber\\
&&\\
Y&=&\frac{1}{a}\wp(w)+b\nonumber\end{eqnarray}
of equation (\ref{eq: BECYeqn}), where again $\zeta(w)$ is the Weierstrass zeta function of (A.5).

\end{example}

\begin{example}
In the main result, equations (2.4), (2.5), $f(x)$ is assumed to have no repeated factors.  However, if repeated factors occur then equation (1.1) can be solved, in fact, in terms of elementary, non-elliptic functions, which we illustrate in the following example.

The equation 
\begin{equation}U'(x)^2+4U(x)^4-2U(x)^2-U(x)/\sqrt{2}=\frac{1}{16},\end{equation}
an example of equation (1.1) with $n=0$, is satisfied by the potential $U(x)$ of a particular \emph{Zakharov-Shabat system}.  $U(x)$, moreover, satisfies a type of static, \emph{modified Novikov-Veselov} (mNV) \emph{equation} \cite{KT}, a point which we return to later.  Here $f(x)=-4x^4+2x^2+x/\sqrt{2}+\frac{1}{16}$ has $(x-x_0)$ as a repeated factor for $x_0\stackrel{def.}{=}-\sqrt{2}/4$, $f(x)=(x-x_0)^2\left[ B_1+B_2(x-x_0)+B_3(x-x_0)^2\right]$ being its finite Taylor expansion about $x_0$ for $2B_1=f''(x_0), 6B_2=f'''(x_0), 24 B_3=f''''(x_0)$.  By the substitution $x=x_0+\frac{1}{u}$,
\begin{equation} I\stackrel{def.}{=}\displaystyle\int \frac{dx}{\sqrt{f(x)}}=\pm \displaystyle\int \frac{du}{\sqrt{B_1 u^2+B_2 u+B_3}}\end{equation}
is an elementary integral (compare the integral in (2.1)) whose evaluation depends on the signs of $B_1$ and the discriminant $\Delta\stackrel{def.}{=}B_2^2-4B_1B_3$.  Actually, as $x_0=-\sqrt{2}/4, B_1=-1<0$ and $\Delta = 16 >0$. 

 In the end, for 
\begin{equation}I = -\frac{1}{\sqrt{-B_1}}arcsin\left( \frac{B_2+2B_1 u}{\sqrt{\Delta}}\right)\end{equation}
and an integration constant $\delta$ we obtain the solution
\begin{equation}U(x)=U_{\mp}(x; \delta) = \frac{\mp sin(x-\delta)}{ 2\sqrt{2}\left[ \sqrt{2}\pm sin(x-\delta)\right]  }\end{equation}
of equation (3.20).  In particular, one has the solution $U_+(x; 0)$ obtained in \cite{KT} (by a quite different method) where among other results the authors there demonstrate invariance of the \emph{Willmore functional} $W$ (the Polyakov extrinsic string action) under NV deformations. 

We note that the ``deformation"
\begin{equation}U(x,t)\stackrel{def.}{=}U_+(x+2t; 0) \stackrel{def.}{=} \frac{sin(x+2t)}{2\sqrt{2}\left[ \sqrt{2} - sin(x+2t) \right]}\end{equation}
of $U_+(x; 0)$ is a solution of the mNV equation
\begin{equation}U_t=U_{xxx}+24U^2 U_x.\end{equation}
The functional $W$ of course is a basic quantity in the study of two-dimensional gravity.

\end{example}




\appendix

\section{}

Given the central importance of the Weierstrass $\wp-$function $\wp(w)$ for the present work we recall briefly, for the reader's convenience, its construction/definition.  As we have indicated in section 2 a more detailed account is available in \cite{Chandra, Greenhill, WW}.  

Let $\omega_1, \omega_2$ be non-zero complex numbers.  Since the imaginary parts of a non-zero complex number $z$ and its reciprocal are related by $Im \ z^{-1}=-\left( Im \ z\right) \left| z \right|^{-2}$, one has that $Im \ \omega_2/\omega_1\neq 0$ if and only if $Im \ \omega_1/\omega_2\neq 0$.  In particular we assume that $Im \ \omega_2/\omega_1>0$, which is equivalent to the assumption $Im \ \omega_1/\omega_2<0$.  The corresponding \emph{lattice} $\mathscr{L}=\mathscr{L}(\omega_1, \omega_2)$ generated by $\omega_1$ and $\omega_2$ is defined to be the set of points $\omega=m\omega_1+n\omega_2$ where $m$ and $n$ vary over the set of whole numbers.  The lattice $\mathscr{L}$ gives rise to the $\wp-$function
\begin{equation}\wp(w)\stackrel{def.}{=}\frac{1}{w^2}+\displaystyle\sum_{\omega\in\mathscr{L}-\{0\}}\left[ \frac{1}{(w-\omega)^2}-\frac{1}{\omega^2}\right]\label{eq: defnwp}\end{equation}
which is also denoted by $\wp(w; \mathscr{L})$, or by $\wp(w; \omega_1, \omega_2)$.  $\wp(w)$ is a meromorphic function, which is doubly periodic with periods $\omega_1, \omega_2$.  Thus, by definition, $\wp(w)$ is an \emph{elliptic} function.  $\wp(w)$ has double poles at $w=\omega\in\mathscr{L}$, and it satisfies the differential equation
\begin{equation}\wp'(w)^2=4\wp(w)^3-g_2(\omega_1, \omega_2)\wp(w)-g_3(\omega_1, \omega_2)\end{equation}
for \emph{invariants}
\begin{equation}
g_2(\omega_1,\omega_2)\stackrel{def.}{=}60\displaystyle\sum_{\omega\in\mathscr{L}-\{0\}} \frac{1}{\omega^4}, \ \ g_3(\omega_1, \omega_2)\stackrel{def.}{=}140 \displaystyle\sum_{\omega\in\mathscr{L}-\{0\}}\frac{1}{\omega^6}\end{equation}
where, moreover,
\begin{equation}g_2(\omega_1, \omega_2)^3-27 g_3(\omega_1, \omega_2)^2\neq 0.\end{equation}

Conversely, it is an amazing fact that if two complex numbers $g_2$ and $g_3$ are given that satisfy the condition $g_2^3-27g_3^2\neq 0$, then there exists a pair of non-zero complex numbers $\omega_1, \omega_2$ with $Im \ \omega_2/\omega_1>0$ such that $g_2(\omega_1, \omega_2)=g_2$ and $g_3(\omega_1, \omega_2)=g_3$, for $ g_2(\omega_1, \omega_2)$ and $ g_3(\omega_1, \omega_2)$ defined in (A.3) with respect to the lattice $\mathscr{L}=\mathscr{L}(\omega_1, \omega_2)$ generated by $\omega_1$ and $\omega_2$.  Thus from $g_2$ and $g_3$ one can also construct the corresponding $\wp-$function $\wp(w; \omega_1, \omega_2)$ (according to definition (A.1)), which in this case we also denote by $\wp(w; g_2, g_3)$ -- as we have so done in the previous sections.

Associated with $\wp(w)$ are the Weierstrass sigma and zeta functions $\sigma(w)$ and $\zeta(w)$, respectively:
\begin{eqnarray}
\zeta'(w)\stackrel{def.}{=}-\wp(w), && \displaystyle\lim_{w\rightarrow 0} \left( \zeta(w)-\frac{1}{w}\right)\stackrel{def.}{=}0,\notag\\
&&\label{eq: defnzetasigma}\\
\notag\frac{\sigma'(w)}{\sigma(w)}\stackrel{def.}{=}\zeta(w), && \displaystyle\lim_{w\rightarrow 0} \frac{\sigma(w)}{w}\stackrel{def.}{=}1.\end{eqnarray}


\end{document}